\documentclass[a4paper,11pt]{article}
\usepackage{multicol}
\usepackage{pos}
\usepackage{subfig}
\usepackage{svg}
\usepackage{hyperref}
\usepackage{graphicx}

\title{Towards more accurate $B_{(s)}\rightarrow\pi(K)$ and $D_{(s)}\rightarrow\pi(K)$ form factors}

\author*[a]{Logan Roberts}
\author[a]{Chris Bouchard}
\author[a]{Olmo Francesconi}
\author[b]{Will Parrott}

\affiliation[a]{University of Glasgow,\\
  University Avenue G12 8QQ, Glasgow, United Kingdom}

\affiliation[b]{York Uiversity, Toronto,\\
4700 Keele St, North York, ON M3J 1P3, Canada}

\emailAdd{l.roberts.1@research.gla.ac.uk}
\emailAdd{chris.bouchard@glasgow.ac.uk}
\emailAdd{olmo.francesconi@glasgow.ac.uk}
\emailAdd{parrott@yorku.ca}

\abstract{We present progress on the calculation of scalar, vector, and tensor form factors for the following meson decays: $B\rightarrow\pi$, $B_s\rightarrow K$, $D\rightarrow\pi$ and $D_s\rightarrow K$. This calculation uses the MILC HISQ gluon field ensembles with HISQ valence quarks. We generate ensembles of correlator data with varying lattice spacings, some as small as 0.044 fm. Some ensembles have a strange-to-light quark mass ratio of 5:1 and others use the physical light quark mass. The fully-relativistic, heavy-HISQ approach is used for the heavy quark, with simulation masses ranging from the charm to near the bottom. This heavy-HISQ approach provides nearly full coverage of the kinematic range. 
}

\FullConference{The 41st International Symposium on Lattice Field Theory (LATTICE2024)\\
 28 July - 3 August 2024\\
Liverpool, UK\\}

\begin{document}
\maketitle

\section{Project Motivations and Specifications}
    The primary aim of this ongoing project is to calculate the vector form factor $f_+(q^2)$ for the $B\rightarrow \pi$ meson decay. This form factor is related to its differential decay rate and the Cabibbo–Kobayashi–Maskawa (CKM) matrix element $|V_{ub}|$ via the following equation \cite{PhysRevD.73.074502}:
        \begin{equation}\label{eq:vub_to_ckm}
            \dfrac{d\Gamma}{dq^2} = \frac{G_F^2}{24\pi^3}p_\pi^3 \times |V_{ub}|^2 \times |f_+(q^2)|^2,
        \end{equation}
    where $G_F$ is the Fermi constant, and $p_\pi$ is the magnitude of the pion's three-momentum in the $B$ meson's rest frame.  This CKM matrix element $V_{ub}$ is similarly present in the $B_s\rightarrow K$ meson decay where the heavy meson contains a strange spectator quark.  
    
     In this project we use the Highly Improved Staggered Quark (HISQ) formalism \cite{HISQ:2006rc} on the MILC $N_f = 2 + 1 + 1$ gluon field ensembles \cite{MILC_2010,MILC_2012}. We follow similar methods found in \cite{Will_towardsB, Will_technical} to perform this new line of form factor calculations.  Our use of the heavy-HISQ approach dictates that meson decays are simulated on the lattice using a generic heavy meson $H$. This generic heavy meson contains a generic heavy quark $h$, whose mass is varied.  In all gluon field ensembles used in this work, the smallest simulated heavy quark's mass is tuned such that it is equal to the physical charm quark mass.  For these reasons, our data sets can be used to calculate form factors for the following four meson decays $B\rightarrow\pi$, $B_s\rightarrow K$, $D\rightarrow\pi$, and $D_s\rightarrow K$.  In the case of $D \rightarrow \pi$ and $D_s \rightarrow K$, the associated CKM matrix element is $V_{cd}$.  Table \ref{tab:lattice_specs} gives a partial description of the current\footnote{We are actively producing data for four more ensembles to use in this work.  These ensembles have coarser lattice spacings with $a \approx 0.12$ fm, $0.15$ fm, two of which include physical light quark masses.} gluon field ensembles used in this project.
    
    The Particle Data Group's (PDG) current listed values for these matrix elements are \cite{ParticleDataGroup:2024cfk}: $|V_{cd}| = 0.221 \pm 0.004$, $|V_{ub}|= (3.82 \pm 0.20)\times 10^{-3}$.  However this given value for $|V_{ub}|$ is a weighted average from both inclusive \textit{and} exclusive $B$ decays.  The PDG lists an exclusive value for $|V_{ub}|$ derived only from lattice QCD and experimental decay rates. Ultimately it is this exclusively determined value of $|V_{ub}|$ we aim to refine through this work.
    
        \begin{table}[b]
        \begin{small}
        \centering
        \begin{tabular}{|c|cccccc|}
        \hline
             Set &  $\approx a$(fm) & $am_l$ &$N_x^3 \times N_t$ & $am_h$ & ${p}_{\pi, K}^\text{max}(\text{MeV})$& $T$\\ \hline \hline
             f-5& 0.09 & 0.0074 & $32^3 \times 96$ & 0.450, 0.55, 0.675, 0.8 & 311 & 15, 18, 21, 24 \\ \hline
             f-phys& 0.09 & 0.00120 & $64^3 \times 96$ & 0.433, 0.555, 0.678, 0.8 & 330 & 15, 18, 21, 24 \\ \hline
             sf-5& 0.06 & 0.0048 & $48^3 \times 144$ & 0.274, 0.5, 0.65, 0.8 & 622 & 22, 25, 28, 31\\ \hline
             sf-phys& 0.06 & 0.0008 & $96^3 \times 192$ & 0.2585, 0.5, 0.65, 0.8 & 648 & 22, 25, 28, 31 \\  \hline
             uf-5& 0.04 & 0.00316 & $64^3 \times 192$ & 0.194, 0.4, 0.6, 0.8 & 583 & 29, 34, 39, 44 \\\hline 
        \end{tabular}
        \end{small}
        \caption{Gluon field ensembles used in this work.  Data listed is, in order: gluon field ensemble set name, approximate lattice spacing, the mass of the light quark in lattice units $am_l$, lattice spacial and temporal extent $N_x$ and $N_t$, the heavy quark masses simulated on each ensemble in lattice units $am_h$, the maximum 3-momentum of the daughter meson momentum $p_{\pi, K}^\text{max}$ simulated on each ensemble in MeV, and the mother-daughter separation widths $T$ simulated for each ensemble.}
        \label{tab:lattice_specs}
        \end{table}
    
\section{Fit Procedure and Refinement}    
    We perform global (one fit per ensemble) correlator fits to two and three point correlators generated on the MILC ensembles.  The resulting fit parameters are then used to calculate form factor results at simulated values of $q^2$.  We then use the modified z-expansion \cite{PhysRevD.79.013008} to extrapolate our form factor results to the continuum limit, physical light and heavy quark masses, and over the entire $q^2$ range.  This fitting procedure makes extensive use of the \textit{gvar}, \textit{lsqfit}, and \textit{corrfitter} \verb|python| libraries, which are documented in \cite{peter_lepage_2024_14210272, peter_lepage_2024_12690493, peter_lepage_2021_5733391}.
    
    \subsection{Correlator and Form Factor Equations}
        We fit two and three point correlators $C_2$ and $C_3$ to equations \ref{eq:2pt_corr_fit} and \ref{eq:3pt_corr_fit}, using $H\rightarrow\pi$ as an example. $H_s \rightarrow K$ correlators are fit similarly.
        \begin{equation} \label{eq:2pt_corr_fit}
            C_2^\pi(t) = \sum_{i=0}^{N_\text{exp}-1} \Bigl[|A^{\pi,n}_i|^2 (e^{-E^{\pi,n}_it} + e^{-E^{\pi,n}_i(N_t-t)}) -(-1)^t|A^{\pi,o}_i|^2 (e^{-E^{\pi,o}_it} + e^{-E^{\pi,o}_i(N_t-t)}\Bigr]
        \end{equation}
        \begin{multline} \label{eq:3pt_corr_fit}
            C_3^{\pi,H}(t,T) =
            \sum_{i,j=0}^{N_\text{exp}-1} \Bigl[A^{\pi,n}_i J_{ij}^{nn} A^{H,n}_j e^{-E^{\pi,n}_it} e^{-E^{H,n}_j(T-t)} 
             \\
             -(-1)^{(T-t)} A^{\pi,n}_i J_{ij}^{no} A^{H,o}_j e^{-E^{\pi,n}_it} e^{-E^{H,o}_j(T-t)} \\
            -(-1)^t A^{\pi,o}_i J_{ij}^{on} A^{H,n}_j e^{-E^{\pi,o}_it} e^{-E^{H,n}_j(T-t)} \\
             +(-1)^T A^{\pi,o}_i J_{ij}^{oo} A^{H,o}_j e^{-E^{\pi,o}_it} e^{-E^{H,o}_j(T-t)}\Bigr]
        \end{multline}
        Here $N_\text{exp}$ is the number of states in the fit, $N_t$ is the temporal lattice length, and $A$, $E$ denote various meson amplitudes and energies.  The superscripts \textit{o} and \textit{n} denote an oscillating or a non-oscillating state respectively.  Finally of the various three point scalar, vector, and tensor current insertion amplitudes $J \in \{S,V,T\}$, $J_{00}^{nn}$ is the ground state three point amplitude we use to calculate the corresponding lattice matrix element,
        \begin{equation}
        \langle \pi |{J_{\text{latt}}}|H\rangle = 2Z_{\text{disc}}\sqrt{M_HE_\pi} \times J_{00}^{nn}.
        \end{equation}
        With the lattice matrix element we then calculate the form factors $f_0(q^2), f_+(q^2), f_T(q^2, \mu)$:
        \begin{align}
        &\text{Scalar:}\quad \langle\pi|{S_{\text{latt}}}|H\rangle =  f_0(q^2) \times
        \frac{M^2_H - M^2_\pi}{m_h - m_l},\\    
        &\text{Vector:}\quad Z_V \langle{\pi}|V_\text{latt}^\mu|{\hat{H}}\rangle = f_+(q^2)\left(p_H^\mu + p_\pi^\mu - \frac{M^2_H - M^2_\pi}{q^2}q^\mu \right) + f_0(q^2)\frac{M^2_H - M^2_\pi}{q^2}q^\mu,\label{eq:form_factor_vector}\\
        &\text{Tensor:}\quad Z_T(\mu)\langle{\hat{\pi}}|{T_\text{latt}^{k0}}|{\hat{H}}\rangle = f_T(q^2, \mu) \times \frac{2iM_Hp_\pi^k}{M^2_H + M^2_\pi}.\label{eq:matrix_element_tensor}
        \end{align}
        In equation \ref{eq:form_factor_vector} the superscript $\mu \in \{0,1,2,3\}$ denotes a direction on the euclidean lattice $(t,x,y,z)$. We choose the spacial lattice direction $\mu = k = 1$ for the tensor matrix element. Our use of local current operators requires local non-Goldstone pseudoscalar mesons in some cases, denoted by the $(\text{\textasciicircum})$ notation. The renormalization terms $Z_\text{disc}$, $Z_V$, and $Z_T(\mu)$ are calculated in \cite{zdisc, Na_2010, koponen2013shapedk, Hatton_2020}, where in this context $\mu$ is a heavy quark mass dependent renormalization scale.
        
    \subsection{Addressing Fitting Obstacles}
        The large amount of correlator data for this project presents a challenge: global correlator fits often exceed wall clock limits.  For $H\rightarrow \pi$, per ensemble, we fit over several variable combinations.  These include four heavy quark masses $am_h$, four spin-taste copies of $H$ from using local current operators, and five "twists" $\theta$ which fix the daughter meson three momentum\footnote{The magnitude of the meson's momentum is equal in all three spacial dimensions, $|p^1|= |p^2|=|p^3|$.} by $\sqrt{3}\pi \times\theta = a{p}_{\pi}\times {N_x}$. Additionally there are four different mother-daughter meson separation widths $T$, and four three point currents: scalar, temporal vector, spacial vector\footnote{We simulate the vector current component in the temporal direction $\mu=0$ and the spacial directions $k = \mu$. The form factor results derived from either choice have complementary uncertainties across the full $q^2$ range \cite{Cooper_2020}.}, and tensor.  There is a similarly sized correlator data set for $H_s \rightarrow K$ with these same variable combinations.  
        
\begin{figure}
            \centering
            \includegraphics[width=\linewidth]{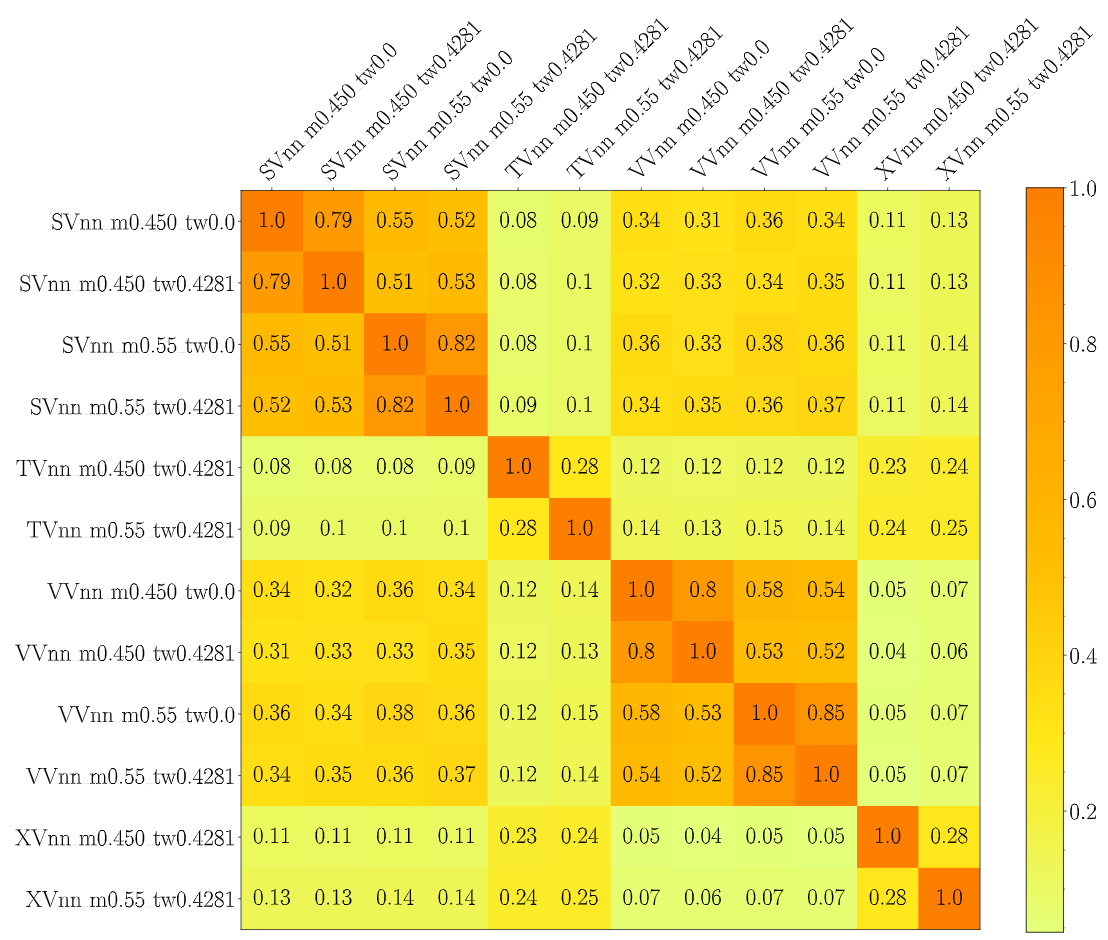}
            \caption{Representative sample three point amplitude correlation matrix for $H\rightarrow\pi$ on the f-5 ensemble.  Each row and column denotes a distinct fit parameter.  The amplitude tags $SVnn$, $TVnn$, $VVnn$, and $XVnn$ denote a scalar, tensor, temporal vector, or spacial vector current respectively.  Parameters shown cover heavy quark masses of $am_h = 0.450, 0.55$, and twists $\theta = 0.0, 0.4281$.  }
            \label{fig:corr-matrix}
        \end{figure}

        It is impractical to perform a simultaneous fit to all the correlators for an ensemble.  We therefore perform a chained fit \cite{peter_lepage_2024_12690493} with links determined by examining the correlations among fit parameters.  Figure \ref{fig:corr-matrix} gives a representative sample of what a fit parameter correlation matrix looks like for a (non-global) parameter subset.  From figures such as this we determine which groups of fit parameters have the largest and smallest correlations and split the global data set into smaller, more manageable subsets.  If two subsets have relatively small statistical correlations with another, we treat them as separate links in the chained fit to all the data.  This preserves correlations between the various subsets of data.
  
        From figure \ref{fig:corr-matrix}, the dominant behavior we observe is that the fit parameters for the four currents split into two subsets.  The scalar current parameters tend to correlate more with the temporal vector current parameters, while the spacial vector current parameters tend to correlate more with the tensor current parameters.  This categorization has correlations that are consistently higher than any alternate grouping of fit parameters, including by different heavy quark masses and twists.

\begin{figure}
        \centering
        \includegraphics[width=0.75\textwidth]{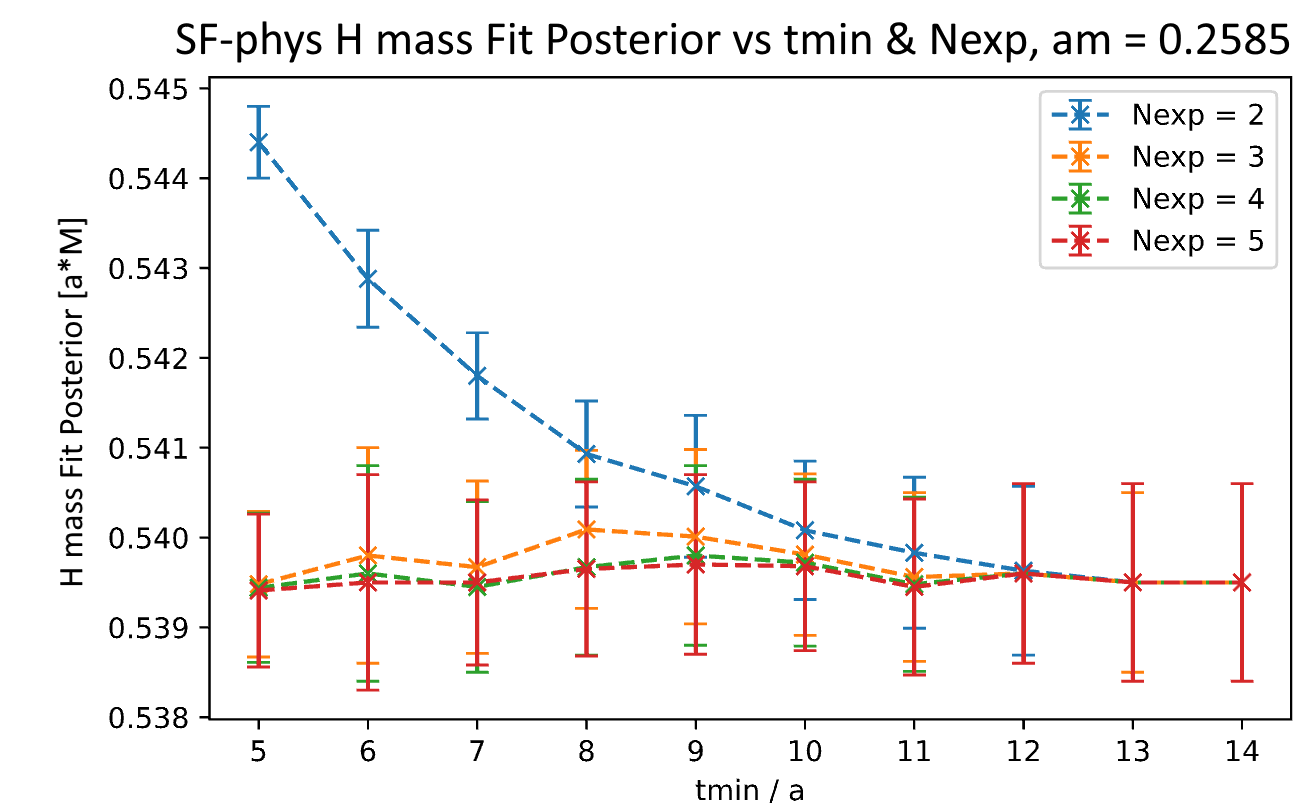}
        \caption{Sample $t_\text{min}$ and $N_\text{exp}$ plot on the sf-phys ensemble.  The mother meson's rest mass, which contains a heavy valence quark $h$ of mass $am_h = 0.2585$, is the fit posterior whose central value and uncertainty is tracked with changing values for $t_\text{min}$ and $N_\text{exp}$.}
        \label{fig:tmin_and_nexp_test}
        \end{figure}

        Another method we can use to reduce time needed is to adjust the number of exponentials in our fit $N_\text{exp}$, and to adjust the $t/a$ domain considered for a given set of correlators.  There are statistical trade-offs in changing these variables; figure \ref{fig:tmin_and_nexp_test} tracks the the change of a sample fit posterior with different values of $N_\text{exp}$ and $t_\text{min}/a$.  In testing we find that a value of $N_\text{exp} = 4$ is more than adequate to capture the higher order behavior in our correlator data in this case.
        
    \subsection{Priors and Bayesian Fitting}
        Following the methodology first outlined in \cite{bayesian}, we provide \textit{a priori}, or prior estimates for posterior fit parameter outputs, which augment the standard $\chi^2$ fitting procedure by equation \ref{eq:chi2_prior_expression}.  
        
        Gaussian priors are specified by a mean and standard deviation (or equivalently, the \textit{central value} and \textit{uncertainty}) of a given energy $E_i$ or amplitude $A_i$, $J_{i}$\cite{bayesian}.   
        Following from equations 5 and 6 in \cite{bayesian}, we augment the normal $\chi^2$ fitting procedure where $\chi^2 \rightarrow \chi^2_\text{aug} = \chi^2 + \chi_{\text{prior}}^2$.  We can denote the prior central value and uncertainty of a given fit parameter $P_i$ as $\Tilde{P}_i + \Tilde{\sigma}_{P_i}$.  With this we define $\chi_{\text{prior}}^2$ as
        \begin{equation}\label{eq:chi2_prior_expression}
            \chi_{\text{prior}}^2 = \sum_i  \frac{(E_i - \Tilde{E}_i)^2}{(\Tilde{\sigma}_{E_i})^2} + \sum_i \frac{(A_i - \Tilde{A}_i)^2}{(\Tilde{\sigma}_{A_i})^2} + \sum_{i}\frac{(J_{i} - \Tilde{J}_{i})^2}{(\Tilde{\sigma}_{J_{i}})^2}.
        \end{equation}
        
    

\begin{figure}
            \centering
             \includegraphics[width=.9\textwidth]{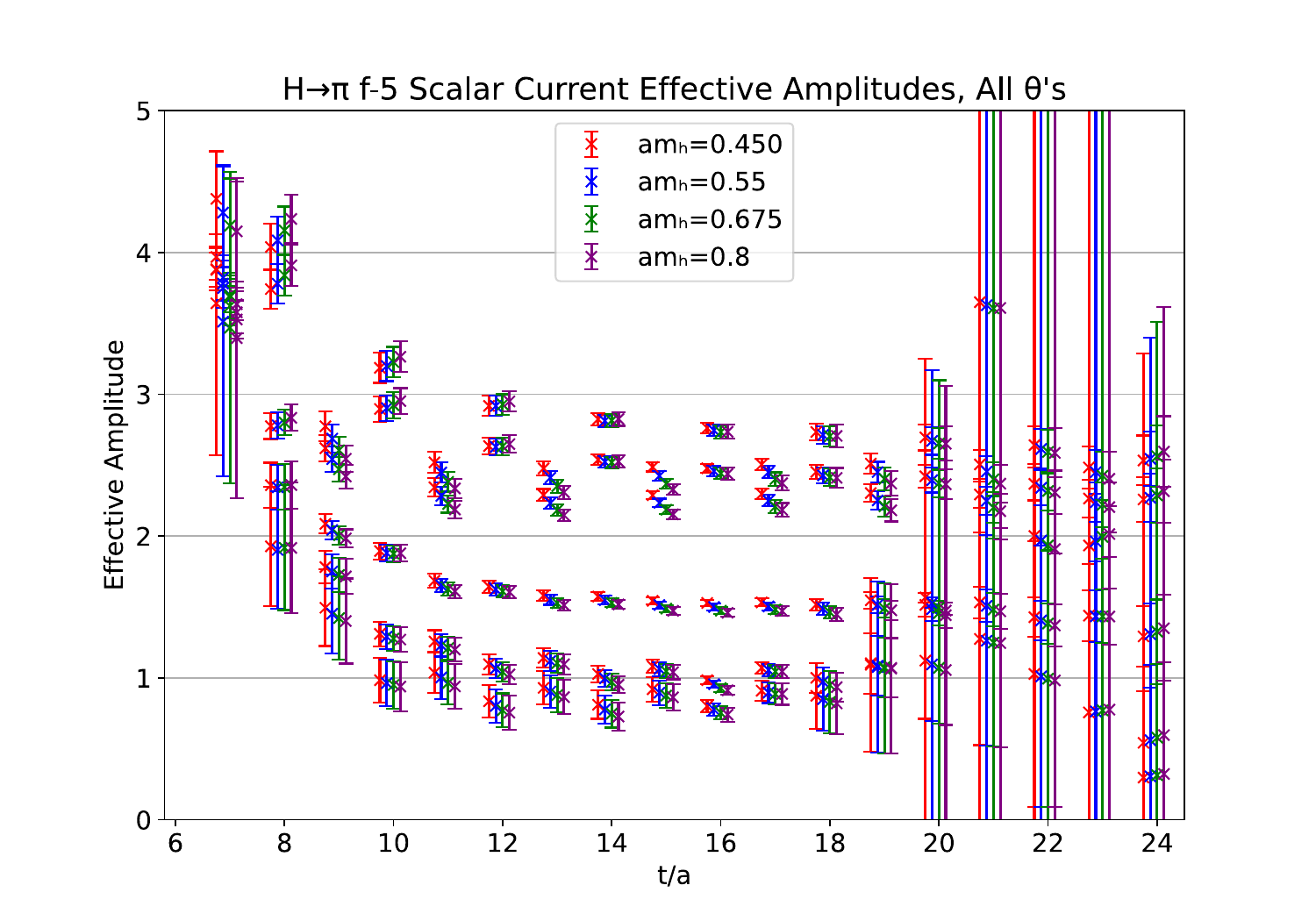}
            \caption{f-5 ensemble $H\rightarrow\pi$ scalar current three-point effective amplitude plot for all mass $am_h$ and twist $\theta$ options.  Different heavy quark mass options for a given $t/a$ are offset along x-axis for visual aid.  The five different twists are represented by the five, typically vertically separated, clusters of $J_{\text{eff}}(t,T)$ for a single $t/a$.}
            \label{fig:priors_3pt_amps}
        \end{figure}

        The practical methods we use to determine the priors for these fit parameters fall into two general categories: manual and automated.  Effective mass $aM_{\text{eff}}(t)$ and amplitude $A_{\text{eff}}(t)$, $J_{\text{eff}}(t,T)$ plots (such as figure \ref{fig:priors_3pt_amps}) can be constructed via equations \ref{eq:2ptam}, \ref{eq:2ptaeff}, and \ref{eq:3ptVeff} from two-point correlators $C_2(t)$ and three-point correlators $C_3(t,T)$.  So long as the correlator data from which these effective mass and amplitude plots are constructed are not too noisy, ground state behavior can be visually (which is to say, manually) differentiated from higher order and oscillating states. 
        \begin{align} \label{eq:2ptam}
        aM_{\text{eff}}(t) &= \frac{1}{2}\cosh^{-1}\left(\frac{C_2(t-2)+C_2(t+2)}{2C_2(t)}\right) \\
        \label{eq:2ptaeff}
        A_{\text{eff}}(t) &= \sqrt{\frac{C_2(t)}{e^{-M_{\text{eff}}t}+e^{-M_{\text{eff}}(N_t-t)}}} \\
        \label{eq:3ptVeff}
        J_{\text{eff}}(t,T) &= C_3(t,T)\frac{\exp[{M_{\text{eff}}^{\pi,K}t +M_{\text{eff}}^{H_{(s)}}(T-t)}]}{A_{\text{eff}}^{\pi,K} \times A_{\text{eff}}^{H_{(s)}}}
        \end{align}

        Meanwhile, some effective mass and amplitude plots are best determined algorithmically.  In such cases where a ground state plateau is sufficiently distinct from non-ground state signal, a function can take the rolling average of adjacent and consecutive $aM_{\text{eff}}(t)$, $A_{\text{eff}}(t)$, or $J_{\text{eff}}(t,T)$ values.  The central value of these ground state mass or amplitude priors can then be automatically determined from a distinct $t/a$ where the change in this rolling average is minimized.
        
\begin{figure}
            \centering
             \includegraphics[width=0.9\textwidth]{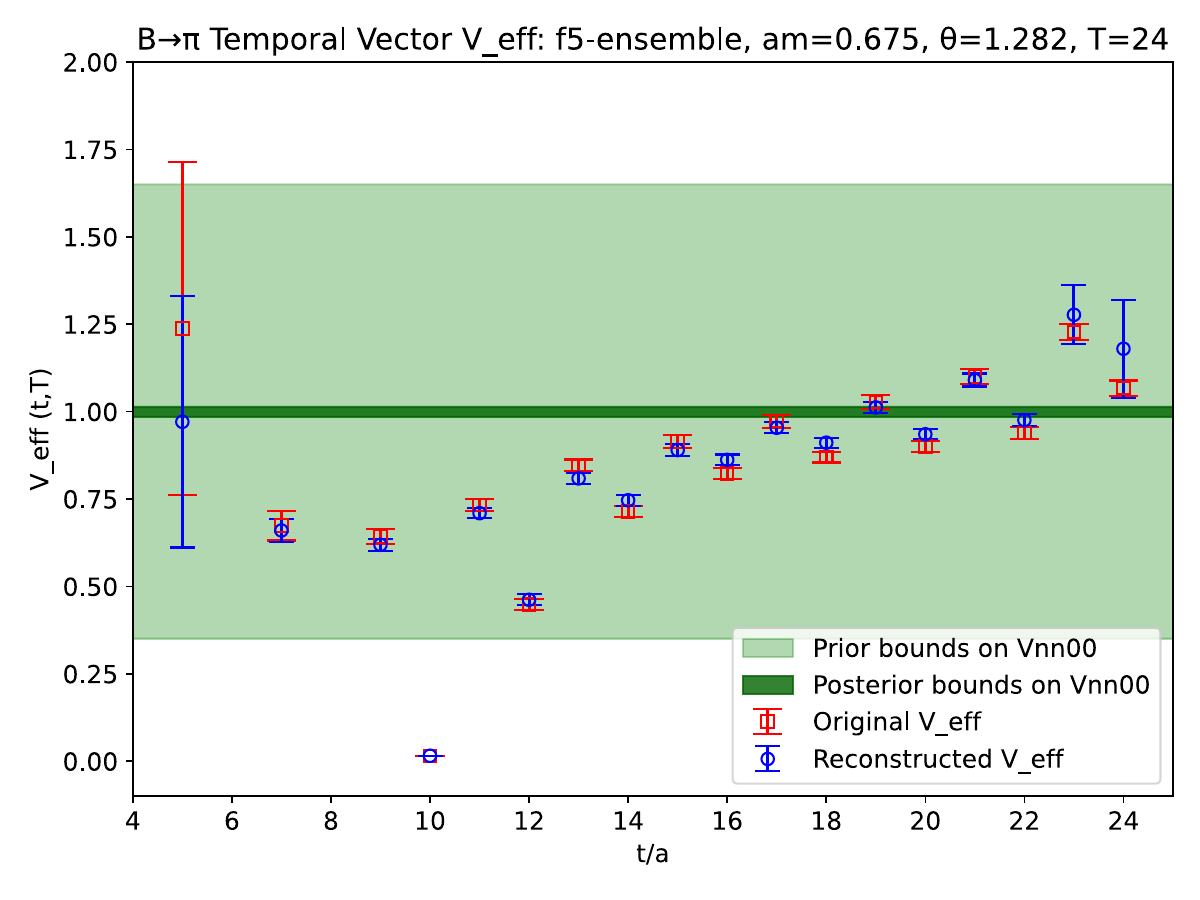}
            \caption{f-5 ensemble $H\rightarrow\pi$ prior vs. posterior sample comparison plot.  Data shown is for a three point temporal vector current with a heavy quark mass of $am_h = 0.675$, a daughter meson twist of $\theta = 1.1282$, and a mother-daughter separation width of $T=24$.  A reconstructed effective amplitude (blue) as a function of $t/a$ is built from reconstructed correlators, which are built from fit two point and three point energies and amplitude posteriors. The original effective amplitude (red) is constructed from original unprocessed correlators.  The light and dark green bands show the prior and posterior bounds on the ground state amplitude $V_{nn}^{00}$ fit parameter associated with the effective amplitude.}
            \label{fig:priors_3pt_comparison}
        \end{figure}

        Irrespective of methodology, we do not use priors as strict constraints on ground state fit parameters. As a guiding principle, we set a prior's uncertainty conservatively, such that its magnitude is no less than ten times its corresponding posterior's uncertainty. Figure \ref{fig:priors_3pt_comparison} gives a typical case of the difference in uncertainty between a prior and its posterior.  We take exception to this principle when we have additional physics information to which we expect the standard model-derived correlator data to adhere. In such cases, physics based algorithms are used to assign values to these priors.  A dispersion relation for example is augmented with known discretization effects and used to set certain pion and kaon priors.

        Finally, in the cases where we cannot easily use the previous methods to determine priors (oscillating excited three point amplitudes for example), we utilize Gaussian Bayes Factor (GBF) optimization. More information on the Bayes factor can be found in \cite{sivia2006data, carlin2010bayes} but in short: a group of prior standard deviations are tuned to maximize a fit's GBF ensuring that the prior neither over-constrains nor under-constrains the fit \cite{bayes_Kass_Rafferty}. A prior is GBF optimized by performing a test fit, and recording the fit's GBF, or more commonly $\log(\text{GBF})$.  Then, the prior's standard deviation is slightly adjusted.  The fit is performed once more and the subsequent change (if any) of its GBF is recorded.  This process is repeated until a clear maximum GBF is resolvable. 

\section{Preliminary Results and Project Outlook}
\begin{figure}
        \centering
        \includegraphics[width = \textwidth]{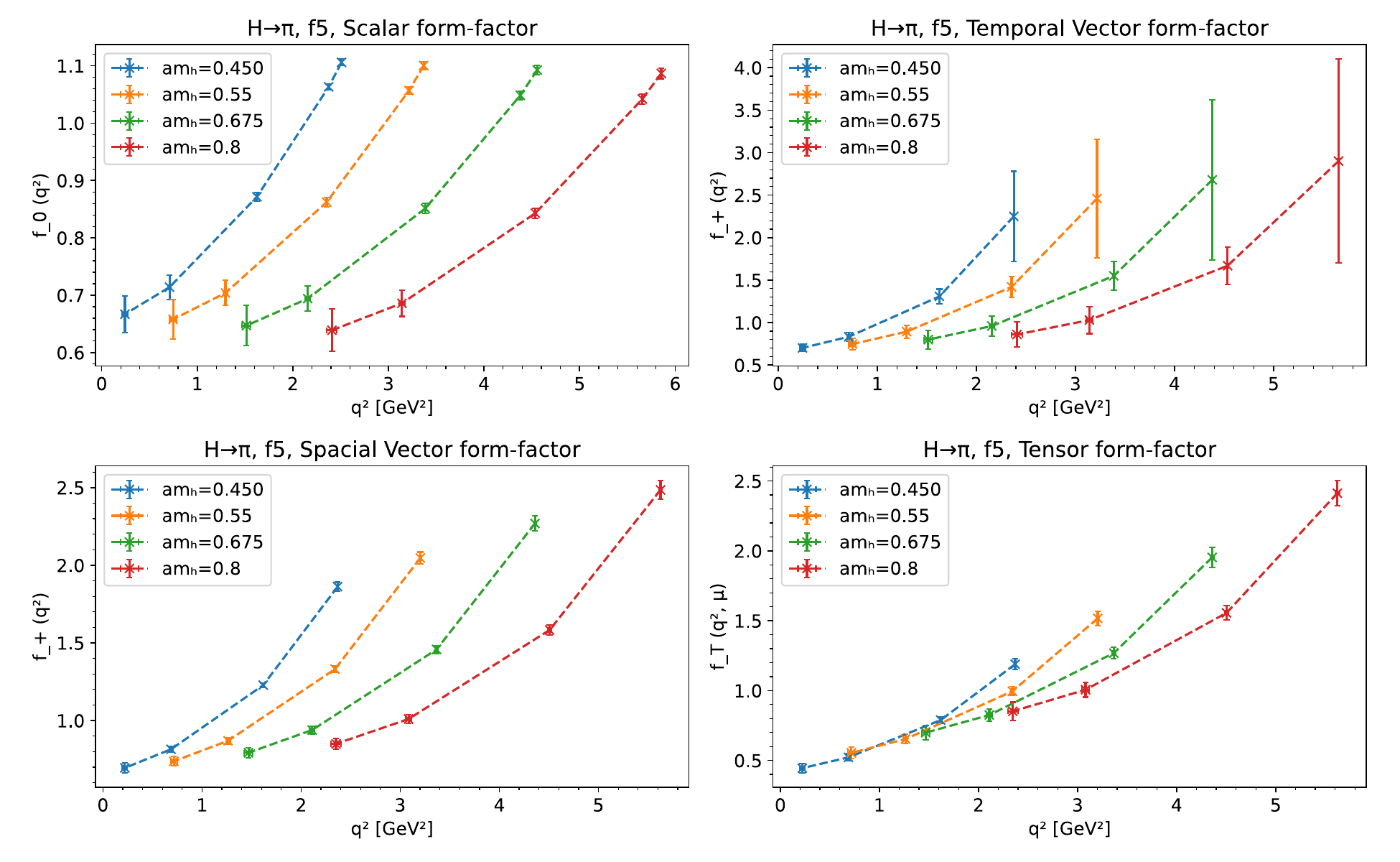}
        \caption{Preliminary $H\rightarrow\pi$ form factor results for all current insertions on the f-5 ensemble.  Results are shown for all heavy quark masses $am_h$ and currents.  Note that $am_h = 0.450$ corresponds to the tuned charm quark mass.}
        \label{fig:prelim_form_factors}
    \end{figure}

    Figure \ref{fig:priors_3pt_comparison} compares a sample three point effective amplitude against its \textit{reconstructed} effective amplitude.  This reconstructed effective amplitude is built using the same procedure as equation \ref{eq:3ptVeff}, but it uses a different three point correlator: one built from fit posteriors via equation \ref{eq:3pt_corr_fit}.  This figure demonstrates how successfully we can capture and model higher order, oscillating, and ground state behavior with our fitting procedure.  Figure \ref{fig:prelim_form_factors} shows preliminary $H\rightarrow\pi$ form factor plots for all four currents on the f-5 ensemble.

    The current focus of this project is to finalize our choice of global ensemble fitting methodology.  Upon the completion of this task, we will convert fit posterior ground state amplitudes to form factor results, and then perform a modified z-expansion to extrapolate to physical quark masses, over the full kinematic range, and to the continuum limit (examples of which can be found in \cite{Will_technical, Cooper_2020, PhysRevD.101.074513, Harrison_2020}). Based on the impressive precision of the $B\rightarrow K$ and $D\rightarrow K$ form factors obtained using the heavy HISQ apprach in \cite{Will_technical}, we anticipate similar improvement here. With these finalized $H_{(s)}\rightarrow\pi (K)$ form factor results we will then estimate values for $|V_{ub}|$ and $|V_{cd}|$, and explore other phenomenological implications of our improved form factors.

\newpage
\providecommand{\href}[2]{#2}\begingroup\raggedright\endgroup

\end{document}